\documentclass[twocolumn,showpacs,aps,preprintnumbers,amsmath,amssymb,superscriptaddress,nofootinbib]{revtex4}

\usepackage{graphicx,subfigure,multirow}
 \usepackage{longtable}
\usepackage{dcolumn}
\usepackage{bm}
\usepackage[all]{xy}
\usepackage{color}
 \usepackage{graphicx}
\usepackage[usenames,dvipsnames]{xcolor}
\usepackage[unicode=true,pdfusetitle,
 bookmarks=true,bookmarksnumbered=false,bookmarksopen=false,citecolor=Turquoise,
 breaklinks=false,pdfborder={0 0 1},backref=false,colorlinks=true,pdfpagemode=FullScreen]
 {hyperref}

\newcommand{\gev}{\textrm{ GeV}}
\newcommand{\GeV}{\textrm{ GeV}}
\makeatletter

\newcommand{\fmslash}[2][0mu]{%
  \mathchoice
    {\fmsl@sh\displaystyle{#1}{#2}}%
    {\fmsl@sh\textstyle{#1}{#2}}%
    {\fmsl@sh\scriptstyle{#1}{#2}}%
    {\fmsl@sh\scriptscriptstyle{#1}{#2}}}
\newcommand{\fmsl@sh}[3]{%
  \m@th\ooalign{$\hfil#1\mkern#2/\hfil$\crcr$#1#3$}}
\makeatother

\newcommand{\lsim}{{\;\raise0.3ex\hbox{$<$\kern-0.75em\raise-1.1ex\hbox{$\sim$}}\;}}
\newcommand{\gsim}{{\;\raise0.3ex\hbox{$>$\kern-0.75em\raise-1.1ex\hbox{$\sim$}}\;}}

\newcommand{\beq}{\begin{equation}}
\newcommand{\eeq}{\end{equation}}
\newcommand{\bea}{\begin{eqnarray}}
\newcommand{\eea}{\end{eqnarray}}
\mathchardef\minus="002D

\newcommand{\amc}{{\sc MadGraph5\textunderscore}a{\sc MC@NLO}}

\newcommand{\mptvec}{{\vec{\fmslash P}_T}}

\newcommand{\met}{{\fmslash E_T}}
\usepackage{color}

\begin{document}
\title{Probing the Triple Higgs Self-Interaction at the Large Hadron Collider}

\author{Jeong Han Kim}
\affiliation{Department of Physics and Astronomy, University of Kansas, Lawrence, KS 66045, USA}
\author{Kyoungchul Kong}
\affiliation{Department of Physics and Astronomy, University of Kansas, Lawrence, KS 66045, USA}
\author{Konstantin T. Matchev}
\affiliation{Physics Department, University of Florida, Gainesville, FL 32611, USA}
\author{Myeonghun Park}
\affiliation{Institute of Convergence Fundamental Studies and School of Liberal Arts, Seoultech, Seoul 01811, Korea}

\date{\today}

\begin{abstract}
We propose a novel kinematic method to expedite the discovery of the double Higgs ($hh$) production in the $\ell^+\ell^- b \bar{b} + \met$ final state. We make full use of recently developed kinematic variables,
as well as the variables {\it Topness} for the dominant background (top quark pair production) and {\it Higgsness} for the signal. 
We obtain a significant increase in sensitivity compared to the previous analyses which used sophisticated algorithms like boosted decision trees or neutral networks. 
The method can be easily generalized to resonant $hh$ production as well as other non-resonant channels.

\end{abstract}



\maketitle

\paragraph*{{\bf Introduction.}}

The discovery of the Higgs boson ($h$) with a mass $m_h=125$ GeV \cite{Aad:2012tfa,Chatrchyan:2012xdj} 
jumpstarted a comprehensive program of the precision measurements of all Higgs couplings.
The current results for the couplings to fermions and gauge bosons \cite{Khachatryan:2016vau}
appear to be in agreement with the Standard Model (SM) predictions. 
However, probing the triple and quartic Higgs self-couplings is notoriously difficult \cite{ATL-PHYS-PUB-2017-001,ATL-PHYS-PUB-2016-024,Kim:2018uty,CMS:2018obr,CMS:2015nat,CMS:2017cwx,Baglio:2012np,Sirunyan:2017guj}.
Yet, the knowledge of those couplings is crucial for understanding the exact mechanism of
electroweak symmetry breaking and the origin of mass in our universe. 

The Higgs self-interaction is parameterized as follows:
\begin{equation}
V = \frac{m_h^2}{2} h^2 + \kappa_3 \lambda_{3}^{\rm SM} v h^3 + \frac{1}{4}\kappa_4  \lambda_{4}^{\rm SM} h^4 \label{eq:Vh} \, ,
\end{equation}
where $\lambda_{3}^{\rm SM} = \lambda_{4}^{\rm SM} = \frac{m_h^2}{2 v^2}$ are the SM values, $\kappa_3$ and $\kappa_4$ parametrize deviations from those, 
and $v \approx 256$ GeV is the Higgs vacuum expectation value. In order to access $\kappa_3$ ($\kappa_4$), one has to measure the process of
double (triple) Higgs boson production at the Large Hadron Collider (LHC) or future colliders, possibly with high luminosity (HL).
Due to the small signal cross-section ($\sigma_{hh}$), it is necessary to combine as many different channels as possible. 
One specific process, $h h \to (b\bar b)(W^\pm W^{\mp})$, has so far been relatively overlooked, due to the 
large SM background cross-section $\sigma_{bknd}\sim 10^5 \sigma_{hh}$, which is predominantly due to
top quark pair production ($t\bar{t}$). In particular, there have been very few studies on the resulting
dilepton final state \cite{Baglio:2012np,CMS:2015nat,Sirunyan:2017guj,CMS:2017cwx,Adhikary:2017jtu}. 
The existing analyses employ sophisticated algorithms (neutral network (NN) \cite{CMS:2015nat}, deep neutral network (DNN) \cite{Sirunyan:2017guj}, 
boosted decision tree (BDT) \cite{Adhikary:2017jtu,CMS:2017cwx}, etc.) to increase the signal sensitivity, but show somewhat pessimistic results,
with a significance no better than $1\sigma$ at the HL-LHC with 3 ab$^{-1}$ luminosity \cite{Sirunyan:2017guj,CMS:2015nat,CMS:2017cwx,Adhikary:2017jtu}.

In this letter, we propose a novel method to enhance the signal significance for $hh$ production in the dilepton channel. The idea is to maximize the use of kinematic information for the dominant background (dilepton $t\bar t$ production).
For this purpose, we utilize the class of kinematic variables which were specifically designed for the dilepton $t\bar{t}$ topology 
\cite{Burns:2008va,Konar:2008ei,Konar:2010ma,Barr:2011xt,Barr:2013tda}. 
In addition, we introduce a discriminator against signal, which we refer to as {\it Topness}, following Ref.~\cite{Graesser:2012qy},
and an analogous discriminator against background, called {\it Higgsness}. 
We do not use matrix elements or any of the above mentioned complicated algorithms. 
The method leads to a surprisingly high significance compared to existing results. 
We will first carry out an analysis for the case of a SM-like Higgs boson, $\kappa_3 = 1$, before extending to non-SM values.

\paragraph*{{\bf Method.}}

Our method relies on two new kinematics functions, Topness and Higgsness, 
which respectively characterize features of $t\bar t$ and $hh$ events, 
and two less commonly used variables, subsystem $M_{T2}$ (or $M_2$) \cite{Lester:1999tx,Burns:2008va,Barr:2011xt} for $t\bar t$ and subsystem $\sqrt{\hat {s}}_{min}$ (or $M_1$) \cite{Konar:2008ei,Konar:2010ma,Barr:2011xt} for $hh$ production.
Topness provides a degree of consistency for a given event to dilepton $t\bar t$ production, in which there are 6 unknowns (the three-momenta of the two neutrinos, $\vec p_{\nu}$ and $\vec p_{\bar\nu}$) and four on-shell constraints, 
$m_t$, $m_{\bar t}$, $m_{W^+}$ and $m_{W^-}$. The neutrino momenta can be fixed by minimizing the following quantity 
\begin{eqnarray}
\chi^2_{ij} &\equiv& \min_{\tiny \mptvec = \vec p_{\nu T} + \vec p_{ \bar\nu T}}  \left [ 
\frac{\left ( m^2_{b_i \ell^+ \nu} - m^2_t \right )^2}{\sigma_t^4} \,   +
\frac{\left ( m^2_{\ell^+ \nu} - m^2_W \right )^2}{\sigma_W^4}  \, \right.  \label{eq:tt} \nonumber \\
&& \hspace*{0.5cm}\left . + \frac{\left ( m^2_{b_j \ell^- \bar \nu} - m^2_t \right )^2}{\sigma_t^4} \, +
\frac{\left ( m^2_{\ell^- \bar\nu} - m^2_W \right )^2}{\sigma_W^4}   \right ]  , 
\end{eqnarray}
subject to the missing transverse momentum constraint, $ \mptvec = \vec p_{\nu T} + \vec p_{ \bar\nu T}$. 
We use \verb|MINUIT| for the minimization in our analysis \cite{James:1975dr}. 
Since there is a twofold ambiguity in the paring of a $b$-quark and a lepton, we define {\it Topness} as the smaller of two $\chi^2$s,
\begin{eqnarray}
T &\equiv&  { min} \left ( \chi^2_{12} \, , \, \chi^2_{21} \right ) \, .
\end{eqnarray}

In double Higgs production, the two $b$-quarks arise from a Higgs decay ($h \to b \bar b$), and therefore their invariant mass $m_{bb}$
can be used as a first cut to enhance the signal sensitivity. 
For the decay of the other Higgs boson, $h \to W^\pm W^{\mp}$, we define {\it Higgsness} as follows: 
\begin{eqnarray}
H &\equiv&    {\rm min} \left [
 \frac{\left ( m^2_{\ell^+\ell^-\nu \bar\nu} - m^2_h \right )^2}{\sigma_{h_\ell}^4}    \right. 
 + \frac{ \left ( m_{\nu  \bar\nu}^2 -  m_{\nu\bar\nu, peak}^2 \right )^2}{ \sigma^4_{\nu}}
   \nonumber \\
 && \hspace*{-0.8cm}   + {min} \left ( 
\frac{\left ( m^2_{\ell^+ \nu } - m^2_W \right )^2}{\sigma_W^4} + 
\frac{\left ( m^2_{\ell^- \bar \nu} - m^2_{W^*, peak} \right )^2}{\sigma_{W^*}^4}  \, , \right. \label{eq:hww}  \\
 && \left.  \left .
\frac{\left ( m^2_{\ell^- \bar \nu} - m^2_W \right )^2}{\sigma_W^4} + 
\frac{\left ( m^2_{\ell^+ \nu} - m^2_{W^*, peak} \right )^2}{\sigma_{W^*}^4}  
\right )  \right ]   \, , \nonumber
\end{eqnarray}
where $m_{W^*}$ is the invariant mass of the lepton-neutrino pair which resulted from the off-shell $W$. 
It satisfies $0 \leq m_{W^*} \leq m_h- m_W$ and the peak of its distribution is at 
$m_{W^*}^{peak} = \frac{1}{\sqrt{3}} \sqrt{ 2 \left ( m_h^2 + m_W^2 \right ) - \sqrt{m_h^4 + 14 m_h^2 m_W^2 + m_W^4}}$.
$m_{\nu\bar\nu}^{peak} = m_{\ell\ell}^{peak} \approx 30$ GeV is the location of the peak in the $\frac{\textrm{d}\sigma}{\textrm{d} m_{\nu\bar\nu}}$ or $\frac{\textrm{d}\sigma}{\textrm{d} m_{\ell\ell}}$ distribution, which is bounded from above by $m_{\nu\bar\nu}^{max} = m_{\ell\ell}^{max} = \sqrt{m_h^2 - m_W^2}$.
The phase space distribution of $\frac{\textrm{d}\sigma}{\textrm{d}m_{\nu \bar\nu}}$ is given by $\frac{\textrm{d}\sigma}{\textrm{d}m_{\nu \bar\nu}} \propto \int\textrm{d}m_{W^*}^2 \lambda^{1/2}(m_h^2,m_W^2, m_{W^*}^2) f(m_{\nu \bar\nu})$, where 
$\lambda(x,y,z)=x^2+y^2+z^2-2xy-2yz-2zx$ is the two-body phase space function and $f(m)$ is the invariant mass distribution of the antler topology \cite{Han:2009ss,Han:2012nr,Han:2012nm,Cho:2012er} with $h\to W W^* \to \ell^+ \ell^- \nu \bar \nu$
\begin{eqnarray}
f(m)\sim \left\{
\begin{array}{l l}
\eta\, m \, , & 0 \leq m \leq e^{-\eta}E, \\ [1mm]
m \ln(E/m)  \, , & e^{-\eta}E \leq m \leq E,
\end{array}\right.  \label{eq:antlerf}
\end{eqnarray}
where now the endpoint $E$ and the parameter $\eta$ are defined in terms of the particle masses as
$E =  \sqrt{ m_W m_{W^*} \, e^{\eta}} $ and $\cosh \eta =\left ( \frac{m_h^2- m_{W}^2 - m_{W^*}^2}{2 m_{W} m_{W^*} } \right )$. 
The actual peak of 30 GeV is slightly less than the result for pure phase space due to a helicity suppression in the $W$-$\ell$-$\nu$ vertex. 

The $\sigma$ values in Eq.\,(\ref{eq:tt}) and Eq.\,(\ref{eq:hww}) are associated with experimental uncertainties and intrinsic particle widths,  but in principle, they can be treated as free parameters and one can tune them using NN, BDT etc. In our numerical study, we use $\sigma_t=5$ GeV, $\sigma_W=5$ GeV, $\sigma_{W^*}=5$ GeV, $\sigma_{h_\ell}=2$ GeV, and $\sigma_\nu = 10 $ GeV. 
The main contribution in Eq.\,(\ref{eq:hww}) comes from the on-shell conditions for the Higgs and the $W$, while the effects of the invariant mass of the two neutrinos and the off-shell $W$ are minor. 

Along with Higgsness and Topness, we adopt the subsystem $\hat{s}_{min}^{(\ell\ell)}$ for $h \to W^\pm W^{*\mp}\to\ell^+\ell^- \nu \bar \nu$ \cite{Konar:2008ei,Konar:2010ma} and the subsystem $M_{T2}$ for the $b \bar b$ system ($M_{T2}^{(b)}$) and the lepton system ($M_{T2}^{(\ell)}$) \cite{Burns:2008va}. 
The variable $ \hat{s}_{min}^{({\rm v})} $ is defined as $ \hat{s}_{min}^{({\rm v})} = m_{{\rm v}}^2 + 2 \left ( \sqrt{ |\vec P_{T}^{\rm v} |^2 + m_{\rm v}^2 }  |\mptvec| - \vec P_{T}^{\rm v} \cdot \mptvec \right )$ \cite{Konar:2008ei,Konar:2010ma,Barr:2011xt}, where $({\rm v})$ represents a set of visible particles under consideration, while $m_{\rm v}$ and $\vec P_{T}^{\rm v}$ are their invariant mass and transverse momentum, respectively. It provides the minimum value of the Mandelstam invariant mass $\hat s$ which is consistent with the observed visible 4-momentum vector.
The $M_{T2}$ is defined as 
$M_{T2} (\tilde m) \equiv \min\left\{\max\left[M_{TP_1}(\vec{p}_{\nu T},\tilde m),\;M_{TP_2} (\vec{p}_{\bar\nu T},\tilde m)\right] \right\}$ where $\tilde m$  is the test mass for the daughter particle and the minimization over the transverse masses of the parent particles $M_{TP_i}$ ($i=1, 2$) is performed over the transverse neutrino momenta $\vec{p}_{\nu T}$ and $\vec{p}_{\bar\nu T}$ subject to the 
$\mptvec $ constraint  \cite{Lester:1999tx,Burns:2008va,Barr:2011xt}. 
  
\vspace*{0.2cm}
\paragraph*{{\bf Analysis.}}

Both the signal and the SM backgrounds were generated by \amc~\cite{Alwall:2014hca} with the default \texttt{NNPDF2.3QED} parton distribution functions~\cite{Ball:2013hta} in leading order QCD accuracy at the $\sqrt{s} = 14$ TeV LHC. %
The default dynamical renormalization and factorization scales were used and off-shell effects for the top quark and $W$ boson are properly included. The double Higgs production cross-section ($\sigma_{hh} = 40.7$ fb) was normalized to 
the next-to-next-to-leading order (NNLO) accuracy in QCD \cite{Grigo:2014jma}. 
Including all relevant branching fractions, we obtain $\sigma_{hh} \cdot 2 \cdot \text{BR}(h \rightarrow b \overline{b}) \cdot \text{BR}(h \rightarrow W W^{*} \rightarrow \ell^+ \ell^- \nu \bar \nu)  =0.648$ fb, where $\ell$ denotes an electron or a muon, including leptons from tau decays. The major $t \overline{t}$ background is normalized to the NNLO QCD cross-section 953.6~pb~\cite{Czakon:2013goa}. 
The next dominant $t \overline{t} h$ background is normalized to the next-to-leading order (NLO) QCD cross-section 611.3~fb~\cite{Dittmaier:2011ti}, while for the $t \overline{t} V$ ($V=W^\pm, Z$) background, we apply an NLO k-factor of 1.54, resulting in 1.71 pb \cite{deFlorian:2016spz}. A conservative NLO k-factor 2 is applied for Drell-Yan backgrounds $\ell \ell b j$ and $\tau \tau b b$, with $j$ representing partons in the five-flavor scheme. Note that a recent study shows ${\rm k}^{NNLO, DY}_{QCD\otimes QED} \approx 1$ \cite{deFlorian:2018wcj}. 
Finally we include the irreducible $jj  \ell\ell \nu\bar \nu$ background from the mixed QCD+EW process with ${\rm k}_{NLO}=2$ 
(denoted as `others' in the rest of this paper).

Both signal and background events are showered and hadronized by \verb|PYTHIA 6| \cite{Sjostrand:2006za}. Jets are clustered with the \verb|FastJet| \cite{Cacciari:2011ma} implementation of the anti-$k_T$ algorithm~\cite{Cacciari:2008gp} with a fixed cone size of $R = 0.4$ for a jet, where the $R$ is the distance from the origin in the ($\eta$, $\phi$) space in terms of the pseudo-rapidity $\eta$ and the azimuthal angle $\phi$. We include semi-realistic detector effects relevant for the HL-LHC based on Ref.~\cite{ATL-PHYS-PUB-2013-004}, where jets, leptons and $\met$ are smeared according to their energies.

We require at least two jets with $p_T> 30\GeV$ and $|\eta^j| < 2.5$, and exactly two leptons with $p_T^\ell  > 20 \gev$, $| \eta^\ell |< 2.5$. 
For the lepton isolation criteria, we impose $p_T^\ell / p^{\Sigma}_T > 0.7$ where $p^{\Sigma}_T$ is the scalar sum of transverse momenta of final state particles (including the lepton itself) within $\Delta R = 0.3$.
We further require that the two leading jets should be $b$-tagged. For $b$-tagging efficiency, we take a flat efficiency of $\epsilon_{b \to b}$ = 0.7 and a mis-tag rate from a $c$-jet (light-jet) as $\epsilon_{c \to b}= 0.2$ ($\epsilon_{j \to b}=0.01$)\,\cite{Sirunyan:2017ezt}.

\begin{figure}[t]
\centering
\includegraphics[width=6cm]{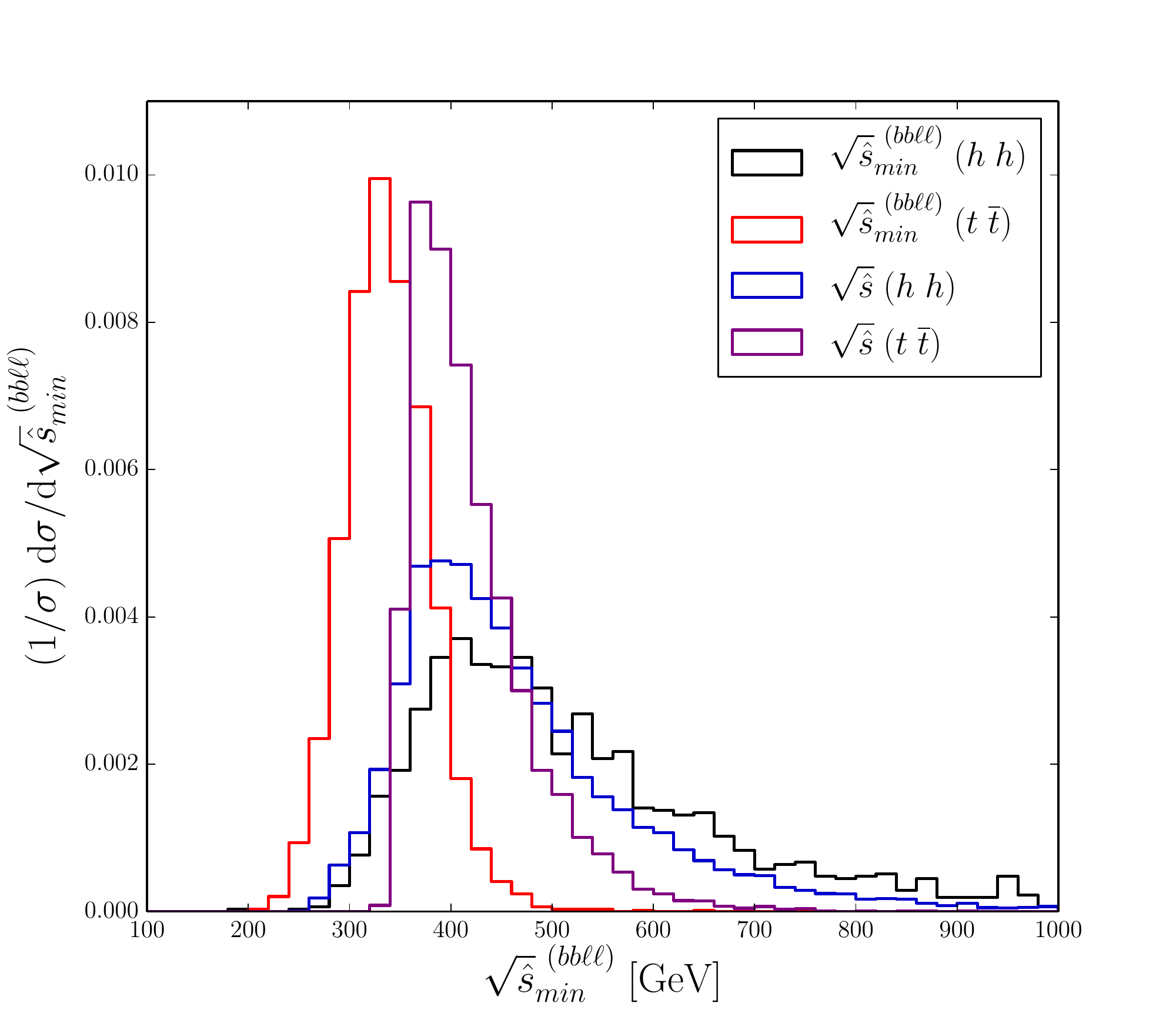}
\caption{\label{fig:sqrts} 
Distributions of $\sqrt{\hat{s}}_{min}^{(bb\ell\ell)}$ and the true $\sqrt{\hat{s}}$ for the case of $hh$ and $t\bar t$ production.
}
\end{figure}
\begin{figure}[t!]
\centering
\includegraphics[width=4.5cm]{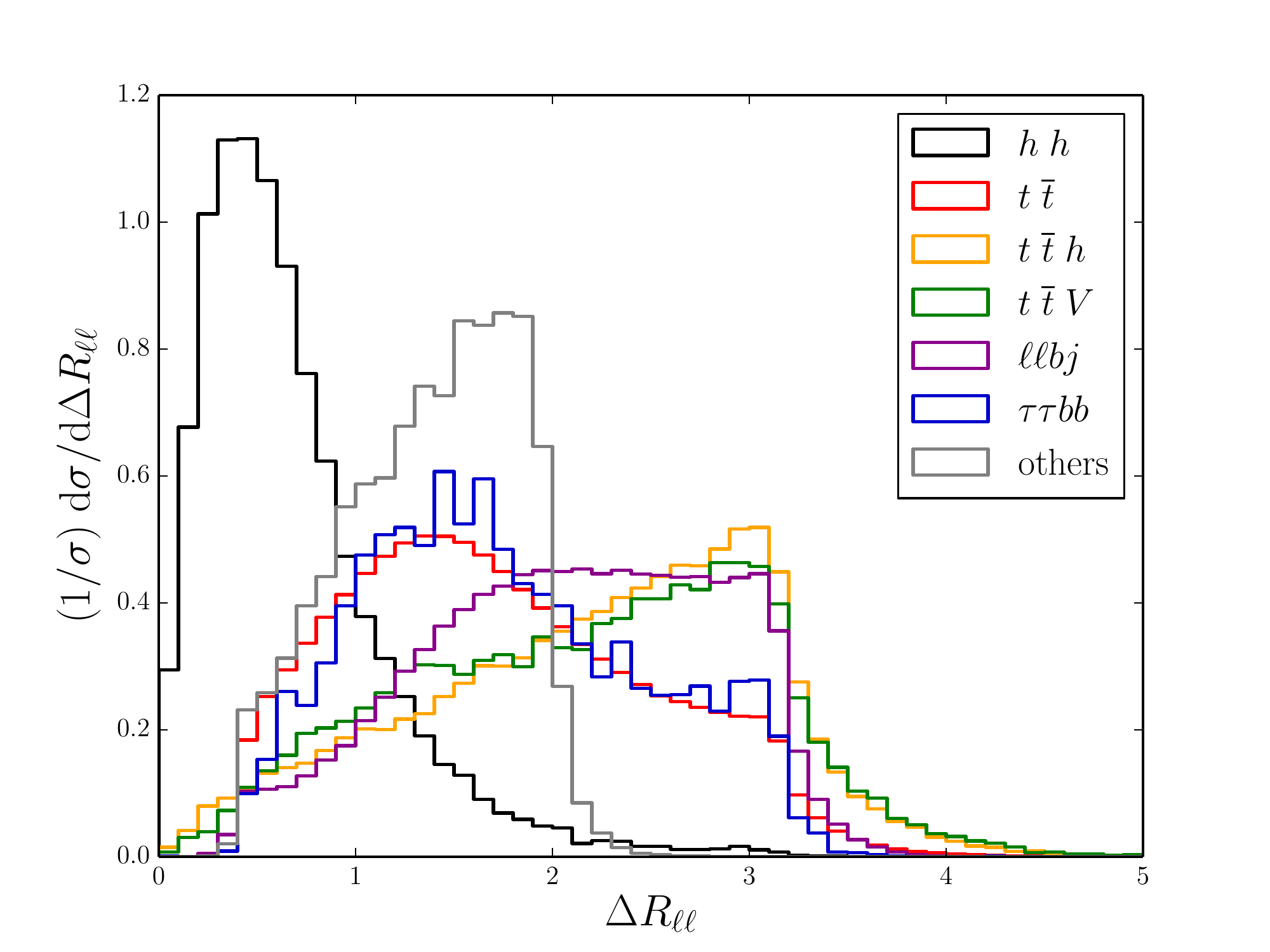} \hspace*{-0.6cm}
\includegraphics[width=4.5cm]{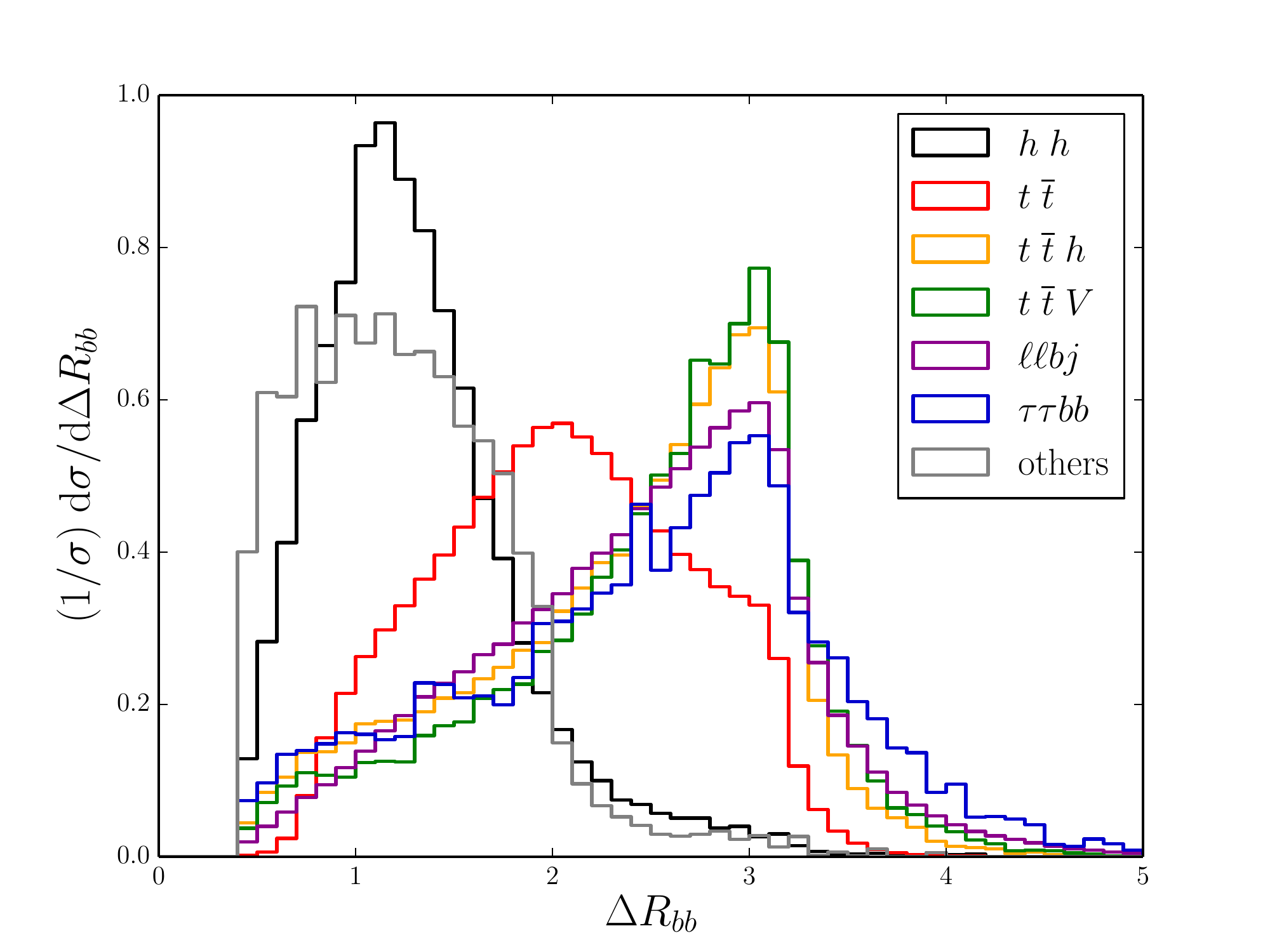} 
\includegraphics[width=4.5cm]{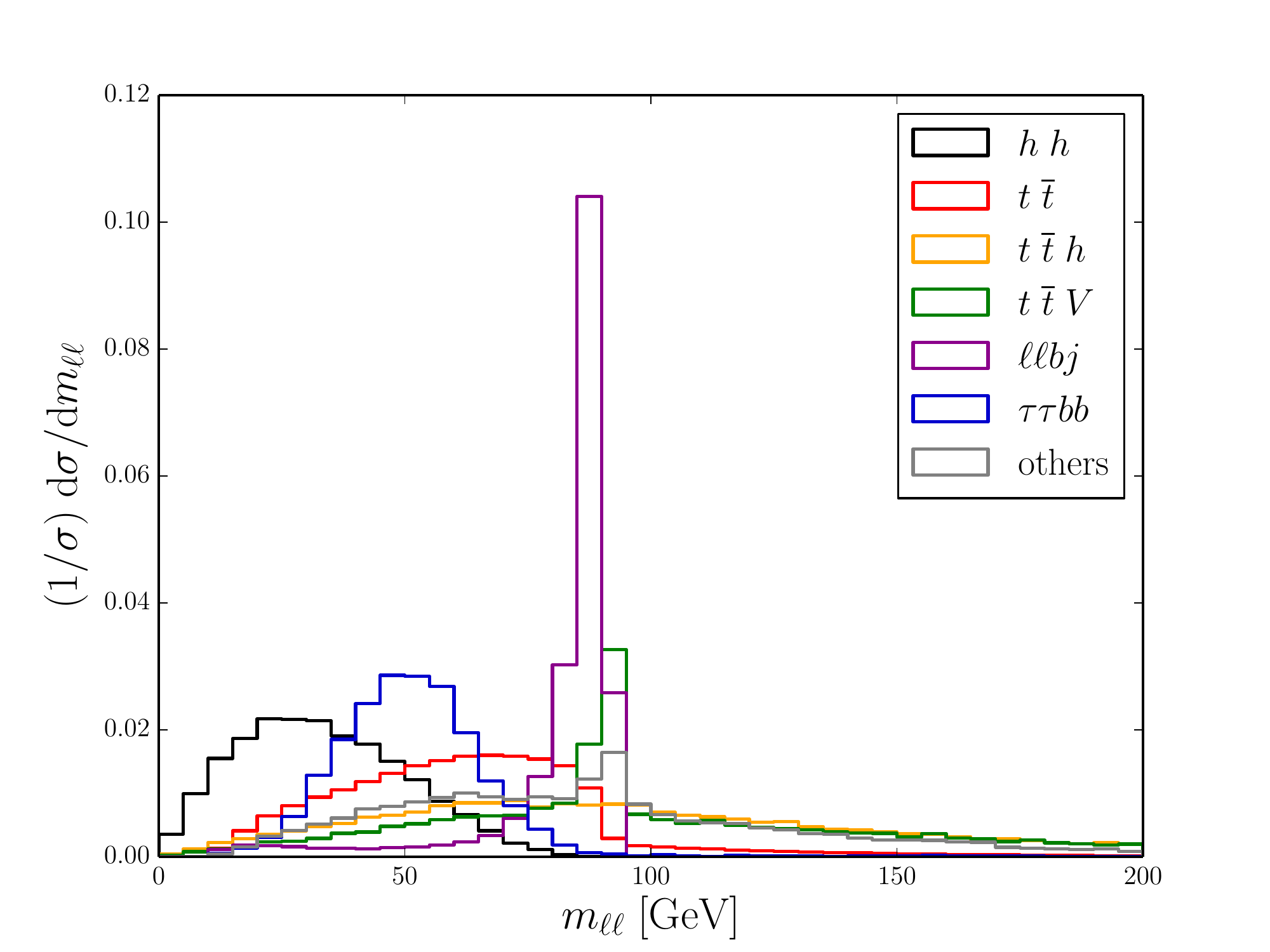}  \hspace*{-0.6cm}
\includegraphics[width=4.5cm]{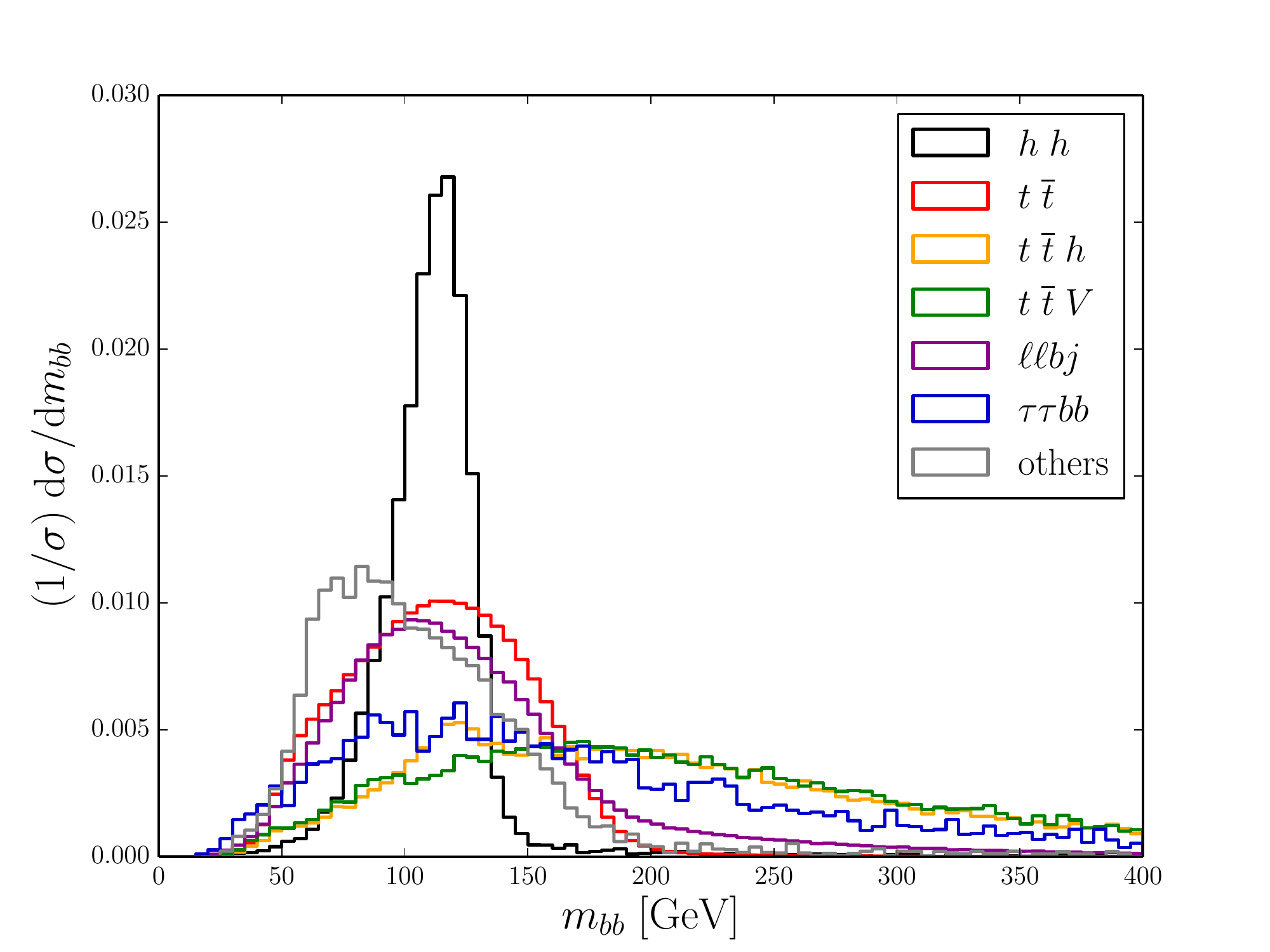} 
\caption{\label{fig:baselinecuts}Distributions of $\Delta R _{\ell\ell}$, $\Delta R_{bb}$, $m_{\ell\ell}$ and $m_{bb}$ after basic generation level cuts.
}
\end{figure}
\begin{figure}[th]
\centering
\includegraphics[width=0.49\linewidth]{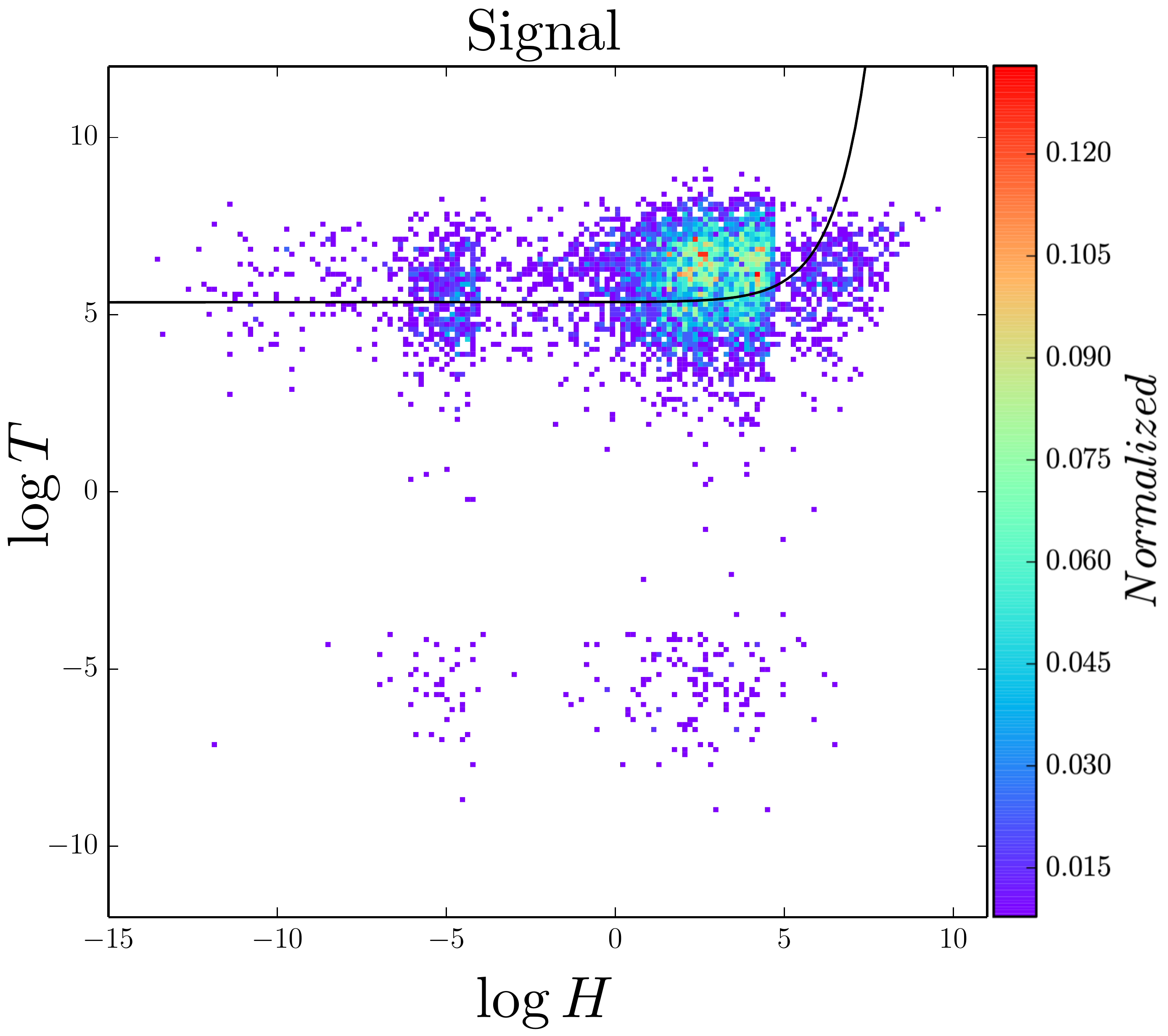}   
\includegraphics[width=0.49\linewidth]{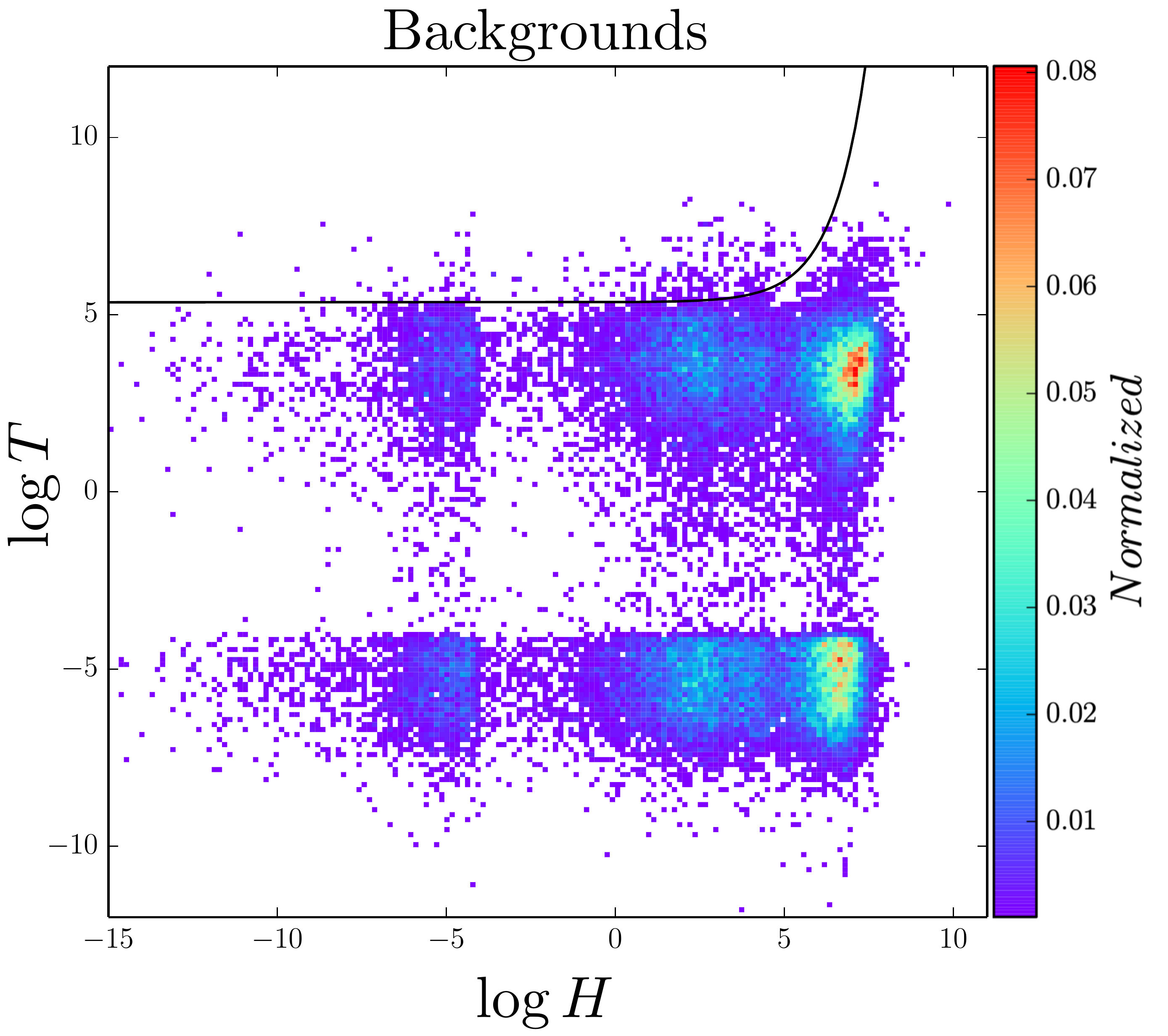} 
\caption{\label{fig:scatter} 
Scatter distribution of ($\log H$, $\log T$) for signal ($hh$) and backgrounds ($t \bar t$, $t\bar t h$, $t \bar t V$, $\ell\ell b j$, $\tau\tau b b$ and others) after loose baseline selection cuts.
The curves are the optimized cuts as in Table\,\ref{tab:Cutflow8}.  
}
\end{figure}

Fig.\,\ref{fig:sqrts} shows distributions of $\sqrt{\hat{s}}_{min}^{(bb\ell\ell)}$ and the true $\sqrt{\hat{s}}$ for $hh$ and $t\bar t$ events. 
First, we observe that $\sqrt{\hat{s}}_{min}^{(bb\ell\ell)}(hh)$ provides a good measure of the true $\sqrt{\hat{s}} (hh)$, while $\sqrt{\hat{s}}_{min}^{(bb\ell\ell)}(t\bar t)$ peaks lower, near the $2 m_t$ threshold. Secondly, 
both $\sqrt{\hat s}(hh)$ and $\sqrt{\hat s }(t\bar t)$ peak at $\sim 400$ GeV. This implies that while the two top quarks are produced near threshold ($2m_t$), the two Higgs bosons are produced well above the corresponding $2m_h$ threshold. 
Consequently, the two top quarks are more or less at rest, while the two Higgs bosons are boosted and their decay products 
tend to be more collimated.
This observation motivates the use of the variables $\Delta R_{\ell\ell}$, $\Delta R_{bb}$, $m_{\ell\ell}$ and $m_{bb}$ for our starting cuts (their individual distributions are shown in Fig.~\ref{fig:baselinecuts}).
These cuts, along with cuts on $\mptvec$ and the lepton transverse momenta $p_T^\ell$ provide our baseline cuts.
Table\,\ref{tab:Cutflow8} lists the corresponding signal and background cross-sections (first row). We then compute 
{\it Topness} and {\it Higgsness} for each event, which provides a pair of likelihoods in the ($\log H$, $\log T$) space.
Our results are shown in Fig.\,\ref{fig:scatter}, where the Higgsness and Topness are chosen as the $x$-axis and $y$-axis, respectively.
The $t\bar t$ events  are expected to be in the bottom right corner (see the right panel), while the $hh$ events are expected to have smaller Higgsness and higher Topness (see the left panel). This motivates the use of a curve in the ($\log H$, $\log T$) space 
as a cut in order to separate signal and background. 
\begin{figure*}[t]
\centering
\includegraphics[width=6.3cm]{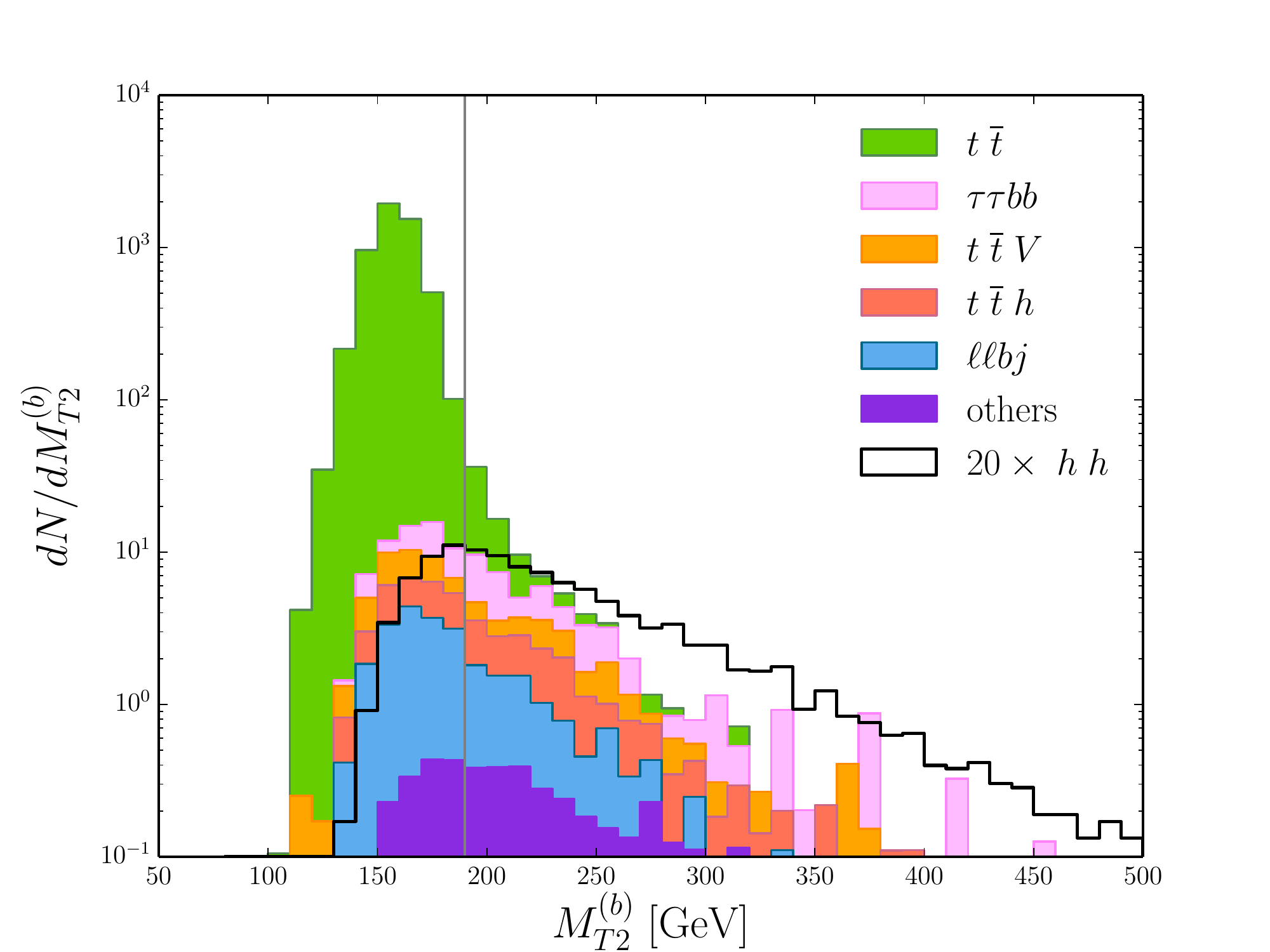} \hspace*{-0.65cm}
\includegraphics[width=6.3cm]{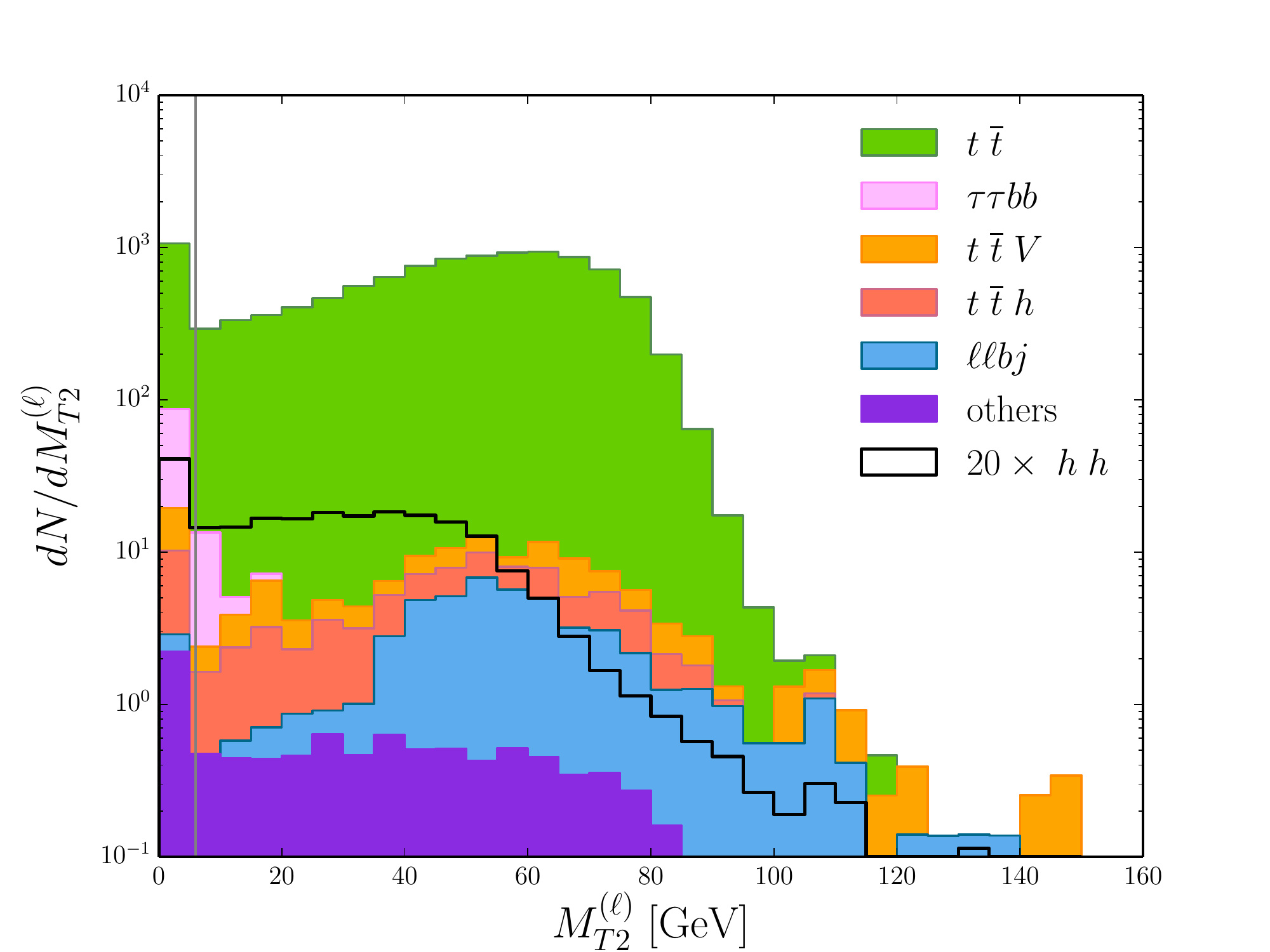}   \hspace*{-0.65cm}
\includegraphics[width=5.95cm]{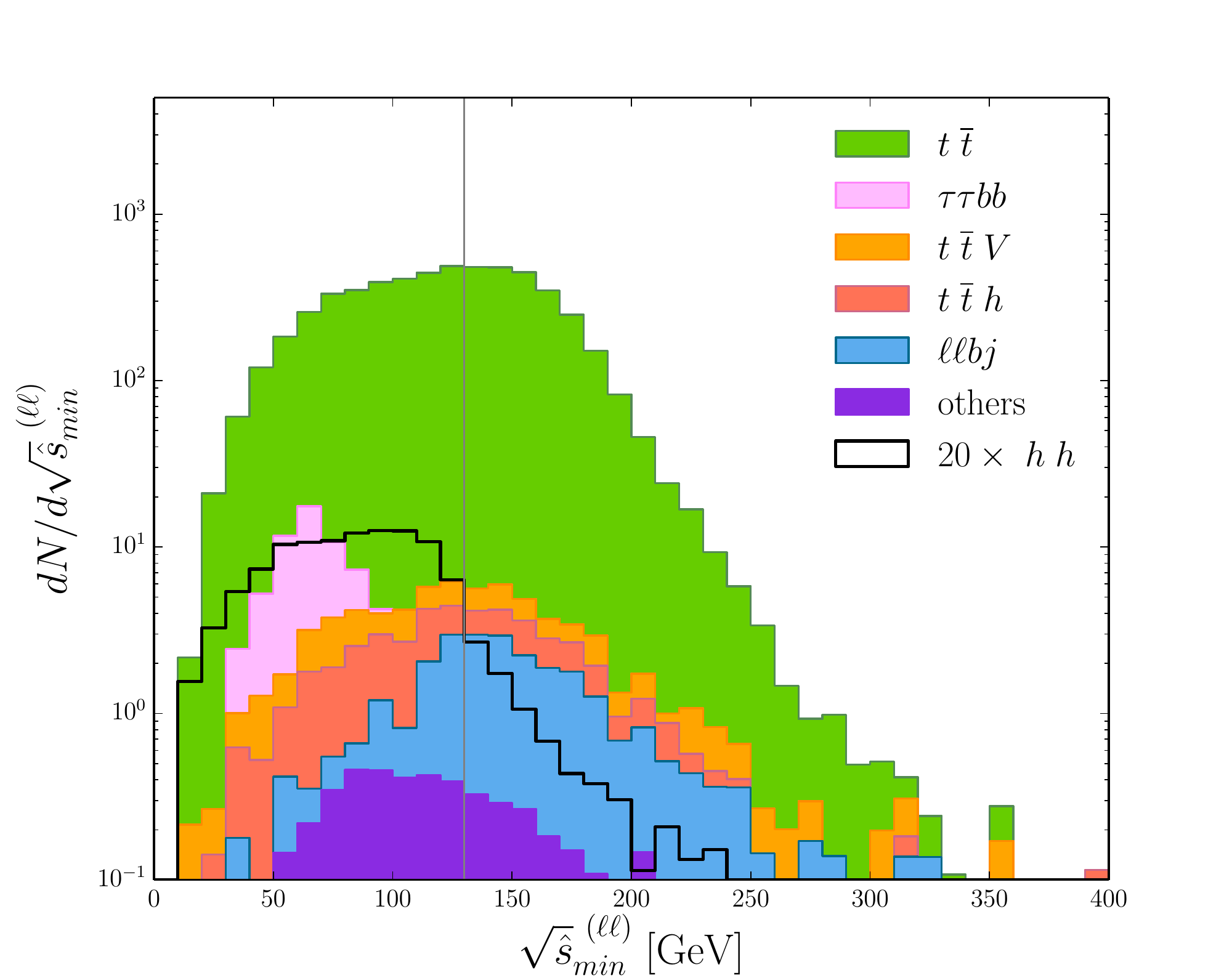} 
\caption{\label{fig:newcuts} 
Distributions for signal ($hh$) and all backgrounds ($t \bar t$, $t\bar t h$, $t \bar t V$, $\ell\ell b j$, $\tau\tau b b$ and others) for $M_{T2}^{(b)}$, $M_{T2}^{(\ell)}$ and $\sqrt{\hat{s}}_{min}^{(\ell\ell)}$ after loose baseline selection cuts. 
The vertical lines at $M_{T2}^{(b)} = 190$ GeV, $M_{T2}^{(\ell)}= 6$ GeV and $\sqrt{\hat{s}}_{min}^{(\ell\ell)}=130$ GeV are the optimized cuts from Table\,\ref{tab:Cutflow8}.
}
\end{figure*}
\begin{table*}[th]
\begin{center}
\setlength{\tabcolsep}{1.1mm}
\renewcommand{\arraystretch}{1.3}
\scalebox{0.98}{
\begin{tabular}{|c||c|c|c|c|c|c|c|c|c|c|}
\hline   
                                                  & Signal                       & $t \overline{t}$       & $t \overline{t} h$     & $t \overline{t} V$     & $\ell \ell bj$                 &$\tau \tau b b$   & {others}  
                                                  & $\sigma$  & {${\rm N_{sig}^{SM}}/{\rm N_{bknd}}$}\\  \hline \hline
{\bf Baseline selections}:  $\met> 20 \gev$,      & \multirow{3}{*}{$0.0124$} &  \multirow{3}{*}{$1.1724$} & \multirow{3}{*}{$0.0297$}  & \multirow{3}{*}{$0.0246$}  &  \multirow{3}{*}{$0.0158$} & \multirow{3}{*}{$0.0379$} & \multirow{3}{*}{{$0.00590$}} & \multirow{3}{*}{{$0.60$}} &  \multirow{3}{*}{{$0.00964$}}  \\   
$p_T^\ell  > 20 \gev$,  $\Delta R_{\ell \ell}<1.0$,  $m_{\ell \ell}<65 \gev$,                             &                                        &                                    &                                        &                                      &                                         &                                      &       &                &     \\  
$\Delta R_{b{b}}<1.3$,  $95<m_{b{b}}<140\gev$  &                                        &                   &                 &                                        &                                      &                                         &                                       &                        &         \\  \hline
${\rm Higgsness} \oplus {\rm Topness} \oplus M_{T2}^{(b)} \oplus  M_{T2}^{(\ell)}  \oplus \sqrt{\hat{s}}_{min}^{(\ell\ell)}$       &  0.0065    &   0.0090      &   0.0073    &  0.0051      &   $ 2.4 \times 10^{-4}$      & 0.0046 & {$0.00185$}         &   {$2.06$}   & {$0.23$}   \\  \hline
\end{tabular}}
\end{center}
\vspace*{-0.3cm}
\caption{
Signal and background cross sections in fb after baseline cuts (first row) and additional cuts (second row), considering ${\rm Higgsness} \oplus {\rm Topness} \oplus M_{T2}^{(b)} \oplus  M_{T2}^{(\ell)}  \oplus \sqrt{\hat{s}}_{min}^{(\ell\ell)}$ requiring {${\rm N_{sig}^{SM}}= 20$}. 
The significance $\sigma$ is calculated using the log-likelihood ratio for a luminosity of 3~$\rm{ab}^{-1}$. 
} 
\label{tab:Cutflow8}
\end{table*}

The second row in Table\,\ref{tab:Cutflow8} lists the signal and background cross-sections after some additional cuts. The last two columns show the corresponding signal significance using the log-likelihood ratio method for a luminosity of 3~$\rm{ab}^{-1}$ and the signal-over-background ratio {${\rm N_{sig}^{SM}/N_{bknd}}$}, respectively. 
Our baseline cuts result in a significance of $0.6$ with {${\rm N_{sig}^{SM}}=37$} and {$\rm N_{bknd}=3841$}, which is in rough agreement with results in literature \cite{CMS:2015nat,Adhikary:2017jtu}. 
We note that one can enhance the signal sensitivity by using ${\rm Higgsness} \oplus {\rm Topness} $ along with $M_{T2}^{(b)}$, $M_{T2}^{(\ell)}$ and $\sqrt{\hat{s}}_{min}^{(\ell\ell)}$. 
The $M_{T2}^{(b)}$ ($M_{T2}^{(\ell)}$) is $M_{T2}$ computed for a subsystem where the two $b$-quarks (leptons) are considered as the visible particles and the two $W$'s ($\nu$'s) as the invisible particles with $\tilde m= m_W$ ($\tilde m= 0$). The $M_{T2}^{(b)}$ for $t\bar t$ events has a kinematic endpoint at $m_t$, while the distributions for the other processes may extend beyond this endpoint, as shown in the left panel of Fig.\,\ref{fig:newcuts}. Similarly, $M_{T2}^{(\ell)}$ has an endpoint at $m_W$ for $t\bar t$ and at $m_\tau$ for $\tau \tau b b$, as shown in the middle panel of Fig.\,\ref{fig:newcuts}. Finally, $\sqrt{\hat{s}}_{min}^{(\ell\ell)}$ in the right panel shows an endpoint at $m_h$ for $hh$ production, while all other backgrounds extend above this point. The cuts on 
Higgsness $\oplus$ Topness, together with the cuts on $M_{T2}^{(b)}$, $M_{T2}^{(\ell)}$ and $\sqrt{\hat{s}}_{min}^{(\ell\ell)}$ increase the significance up to $\sigma=2.1$ and {${\rm N_{sig}^{SM}/N_{bknd}}=0.25$}, keeping {${\rm N_{sig}^{SM}}=20$}. 
The curve in the two scatter plots of Fig.~\ref{fig:scatter} and the vertical lines at $M_{T2}^{(b)} = 190$ GeV, $M_{T2}^{(\ell)}= 6$ GeV and $\sqrt{\hat{s}}_{min}^{(\ell\ell)}=130$ GeV in Fig.~\ref{fig:newcuts} represent the optimized cuts from the second row of Table\,\ref{tab:Cutflow8}.

\begin{figure}[b]
\centering
\includegraphics[width=7.cm, height=5.5cm]{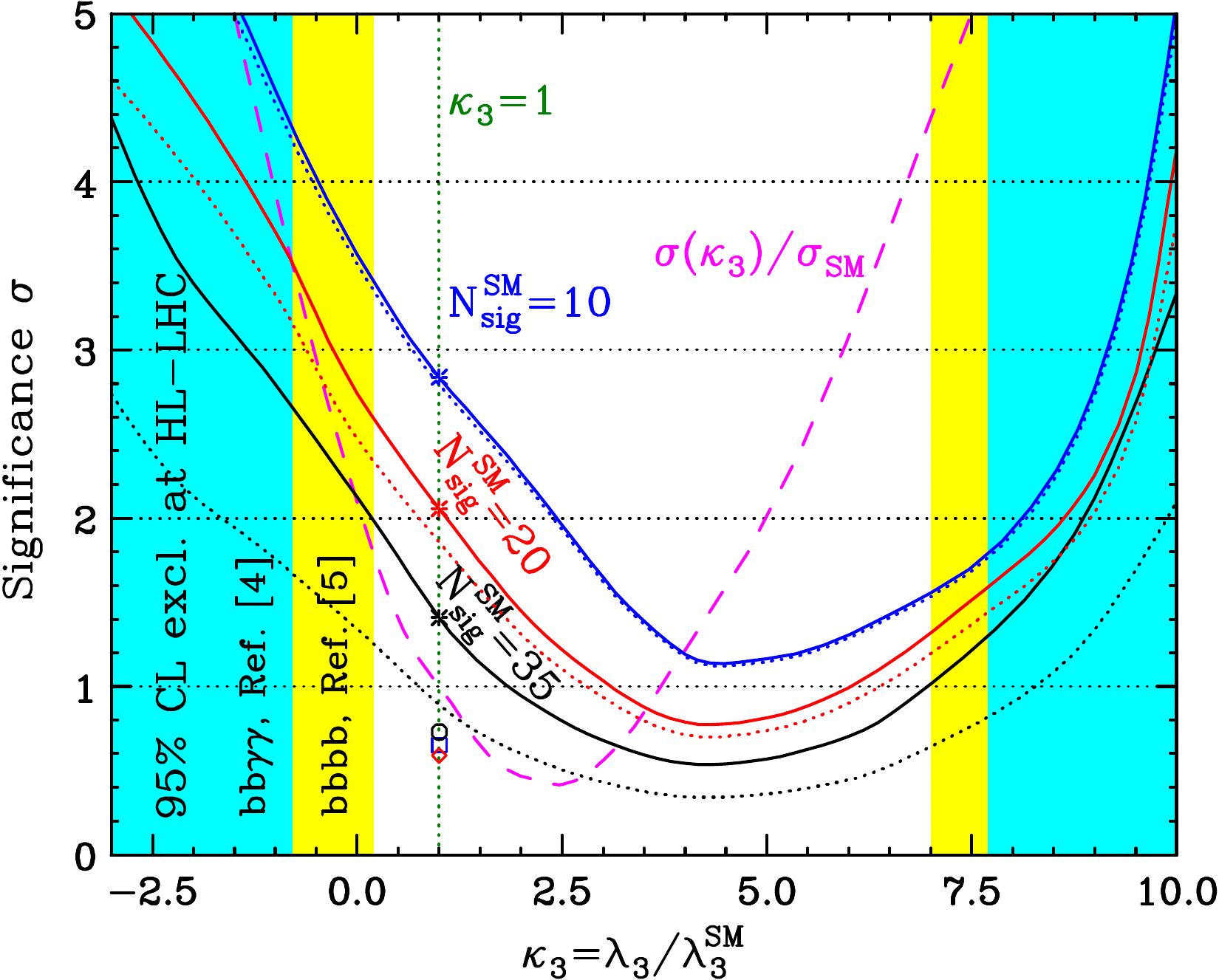}
\vspace*{-0.2cm}
\caption{\label{fig:kappa3} 
Expected significance $\sigma$ at the 14 TeV LHC with 3 ab$^{-1}$ as a function of the triple Higgs coupling $\kappa_3$.
We obtain each curve by applying the same set of cuts optimized for the SM point ($\kappa_3=1$) to non-SM points ($\kappa_3 \neq 1$) for ${\rm N_{sig}^{SM}}=35$ in black, ${\rm N_{sig}^{SM}}=20$ in red and ${\rm N_{sig}^{SM}}=10$ in blue.
The yellow (cyan) shade represents 95\% CL exclusion of $\kappa_3$ in the $bb\gamma\gamma$ channel \cite{ATL-PHYS-PUB-2017-001} ($bbbb$ channel \cite{ATL-PHYS-PUB-2016-024}) at the HL-LHC. The three marks 
{\color{red}$\Diamond$}, {\color{black}$\bigcirc$} and {\color{blue}$\square$} display the signal significance 
using CMS-NN \cite{CMS:2015nat}, CMS-BDT \cite{CMS:2017cwx} and BDT \cite{Adhikary:2017jtu}, respectively.
The dotted curves represent the significance including $5\%$ systematic uncertainty.
}
\end{figure}

A higher significance of 3.0$\sigma$ can be obtained by imposing slightly tighter baseline cuts ($\Delta R_{\ell \ell}<0.48$, $m_{\ell \ell}<60 \gev$, $\Delta R_{b{b}}<1.1$, and $95<m_{b{b}}<140\gev$) and reoptimizing the additional cuts for {${\rm N_{sig}^{SM}}=10$} and {${\rm N_{sig}^{SM}/N_{bknd}}\sim 1.2$}.
Conversely, we can obtain more signal events with slightly looser baseline cuts. For instance, {${\rm N_{sig}^{SM}}=35$} with {${\rm N_{sig}^{SM}/N_{bknd}}=0.061$} is easily obtained with loose baseline cuts ($\Delta R_{\ell \ell}<1.6$, $m_{\ell \ell}<75 \gev$, $\Delta R_{b{b}}<1.5$, and $90<m_{b{b}}<140\gev$) and optimization of ${\rm Higgsness} \oplus {\rm Topness}$, which gives a significance of 1.4$\sigma$.  
This can be compared to existing results with a similar number of signal events but a lower significance of $\sim$0.7 \cite{CMS:2015nat,Adhikary:2017jtu}.

Finally, we perform a similar analysis for different values of $\kappa_3$, and show the expected signal significance as a function of $\kappa_3$ in Fig.\,\ref{fig:kappa3}. We obtain each curve by applying the same set of cuts optimized for the SM point ($\kappa_3=1$) to non-SM points ($\kappa_3 \neq 1$) for ${\rm N_{sig}^{SM}}=35$ in black, ${\rm N_{sig}^{SM}}=20$ in red and ${\rm N_{sig}^{SM}}=10$ in blue. 
The red curve (middle) is the result of Table\,\ref{tab:Cutflow8} for {${\rm N_{sig}^{SM}}=20$}, while the other two curves correspond to tighter (blue) and looser (black) baseline selections for {${\rm N_{sig}^{SM}}=35$} and {${\rm N_{sig}^{SM}}=10$}, respectively.
One can easily compute the number of signal events for $\kappa_3\neq 1$, using the ratio of cross sections of $\sigma(\kappa_3 \neq 1)$ to $\sigma_{SM}=\sigma(\kappa_3 = 1)$ (dashed, magenta) \cite{Baglio:2012np} and ${\rm N_{sig}^{SM}}$ for each SM point.

At 95\% confidence level (CL), the Higgs boson self-coupling is expected to be constrained to $-0.8 < \kappa_3 < 7.7$ ($0.2 < \kappa_3 < 7.0$) from the $bb\gamma\gamma$ final state \cite{ATL-PHYS-PUB-2017-001} ($bbbb$ \cite{ATL-PHYS-PUB-2016-024}) at the HL-LHC, which are shown as yellow and cyan shade, respectively 
(see Ref. \cite{Kling:2016lay} for an improved projection in the $bb\gamma\gamma$ channel.). 
The three symbols {\color{red}$\Diamond$}, {\color{black}$\bigcirc$} and {\color{blue}$\square$} represent the significance $\sigma=0.6$ using CMS-NN \cite{CMS:2015nat}, $\sigma=0.73$ using CMS-BDT \cite{CMS:2017cwx} and $\sigma=0.62$ using BDT \cite{Adhikary:2017jtu}, respectively. Our method clearly shows better performance for both signal sensitivity and exclusion.

\vspace*{0.1cm}
\paragraph*{{\bf Discussion.}}

We have investigated the discovery potential of non-resonant double Higgs production in the final state with $b \bar b \ell^+\ell^- + \met$.
We have shown that the signal sensitivity can be significantly improved by using suitable kinematic variables, without relying on sophisticated methods. Our only assumptions are event topologies of signal and backgrounds with appropriate mass shell conditions. Yet, our final signal significance is higher than existing results in literature at least by a factor of 2, due to to correct use of kinematics for the corresponding event topologies. One could even make further improvement by optimizing cut values and parameters in the discussed method. 

Among various sources of systematic uncertainties, a current CMS study \cite{Sirunyan:2017guj} lists the QCD scale uncertainty (13\%) and the uncertainty in the $t\bar t$ cross-section (5\%), as the dominant ones, and we estimate the final systematic uncertainty to be about 15\% at the 13 TeV LHC. It may be reduced to $\sim$5\% at the HL-LHC \cite{CMS:2015nat}, whose effects are shown in the dotted curves in Fig.\,\ref{fig:kappa3}. 
Our result including the systematic uncertainty still remains very optimistic, given that {${\rm N_{sig}^{SM}/N_{bknd}}$} is large.
Finally, pileup effects may be mitigated by installing new precision timing detectors \cite{MIPtiming} or by analysis techniques using substructure \cite{Asquith:2018igt,Soyez:2018opl,ATLAS:2012am,Bertolini:2014bba} or machine learning \cite{Komiske:2017ubm,Martinez:2018fwc}. In particular, the charged leptons are free of pileup contamination, which further motivates the study of our final state.

The current observed upper limit on the SM $hh \to b \bar b V V \to b \bar b \ell^+ \ell^- \nu \bar \nu$ cross section is found to be 72 fb by CMS using DNN with 36 fb$^{-1}$ of luminosity \cite{Sirunyan:2017guj}. We believe our method can be used in order to improve the signal sensitivity already at the 13 TeV LHC.

\section*{Acknowledgments}
We thank Minho Kim for helpful discussion on BDT and the anonymous referee for constructive comments. 
KK thanks the Aspen Center for Physics for hospitality during the completion of this work, supported in part by National Science Foundation grant PHY-1607611. This work is supported in part by United States Department of Energy (DE-SC0010296, DE-SC0017988 and DE-SC0017965) and Korea NRF-2018R1C1B6006572.



\begin{thebibliography}{999}


\bibitem{Aad:2012tfa} 
  G.~Aad {\it et al.} [ATLAS Collaboration],
  Phys.\ Lett.\ B {\bf 716}, 1 (2012)
  doi:10.1016/j.physletb.2012.08.020
  [arXiv:1207.7214 [hep-ex]].
  
\bibitem{Chatrchyan:2012xdj} 
  S.~Chatrchyan {\it et al.} [CMS Collaboration],
  Phys.\ Lett.\ B {\bf 716}, 30 (2012)
  doi:10.1016/j.physletb.2012.08.021
  [arXiv:1207.7235 [hep-ex]].
  
  
\bibitem{Khachatryan:2016vau} 
  G.~Aad {\it et al.} [ATLAS and CMS Collaborations],
  JHEP {\bf 1608}, 045 (2016)
  doi:10.1007/JHEP08(2016)045
  [arXiv:1606.02266 [hep-ex]].
  



\bibitem{ATL-PHYS-PUB-2017-001} 
  ATLAS Collaboration, 
  ATL-PHYS-PUB-2017-001.

\bibitem{ATL-PHYS-PUB-2016-024} 
  ATLAS Collaboration, 
  ATL-PHYS-PUB-2016-024.

\bibitem{Kim:2018uty} 
  J.~H.~Kim, Y.~Sakaki and M.~Son,
  Phys.\ Rev.\ D {\bf 98}, no. 1, 015016 (2018)
  doi:10.1103/PhysRevD.98.015016
  [arXiv:1801.06093 [hep-ph]].


\bibitem{CMS:2018obr} 
  CMS Collaboration, 
  CMS-PAS-HIG-17-030.


\bibitem{Baglio:2012np} 
  J.~Baglio, A.~Djouadi, R.~GrÃ¶ber, M.~M.~MÃŒhlleitner, J.~Quevillon and M.~Spira,
  JHEP {\bf 1304}, 151 (2013)
  doi:10.1007/JHEP04(2013)151
  [arXiv:1212.5581 [hep-ph]].


\bibitem{CMS:2015nat} 
  CMS Collaboration,
  CMS-PAS-FTR-15-002.
  
  
\bibitem{Sirunyan:2017guj} 
  A.~M.~Sirunyan {\it et al.} [CMS Collaboration],
  JHEP {\bf 1801}, 054 (2018)
  doi:10.1007/JHEP01(2018)054
  [arXiv:1708.04188 [hep-ex]].


\bibitem{CMS:2017cwx} 
  CMS Collaboration, 
  CMS-PAS-FTR-16-002.
  





\bibitem{Adhikary:2017jtu} 
  A.~Adhikary, S.~Banerjee, R.~K.~Barman, B.~Bhattacherjee and S.~Niyogi,
  arXiv:1712.05346 [hep-ph].


\bibitem{Burns:2008va} 
  M.~Burns, K.~Kong, K.~T.~Matchev and M.~Park,
  JHEP {\bf 0903}, 143 (2009)
  doi:10.1088/1126-6708/2009/03/143
  [arXiv:0810.5576 [hep-ph]].
  
  
  
\bibitem{Konar:2010ma} 
  P.~Konar, K.~Kong, K.~T.~Matchev and M.~Park,
  JHEP {\bf 1106}, 041 (2011)
  doi:10.1007/JHEP06(2011)041
  [arXiv:1006.0653 [hep-ph]].
\bibitem{Konar:2008ei} 
  P.~Konar, K.~Kong and K.~T.~Matchev,
  JHEP {\bf 0903}, 085 (2009)
  doi:10.1088/1126-6708/2009/03/085
  [arXiv:0812.1042 [hep-ph]].

\bibitem{Barr:2011xt} 
  A.~J.~Barr, T.~J.~Khoo, P.~Konar, K.~Kong, C.~G.~Lester, K.~T.~Matchev and M.~Park,
  Phys.\ Rev.\ D {\bf 84}, 095031 (2011)
  doi:10.1103/PhysRevD.84.095031
  [arXiv:1105.2977 [hep-ph]].


\bibitem{Barr:2013tda} 
  A.~J.~Barr, M.~J.~Dolan, C.~Englert and M.~Spannowsky,
  Phys.\ Lett.\ B {\bf 728}, 308 (2014)
  doi:10.1016/j.physletb.2013.12.011
  [arXiv:1309.6318 [hep-ph]].
  

  
\bibitem{Graesser:2012qy} 
  M.~L.~Graesser and J.~Shelton,
  Phys.\ Rev.\ Lett.\  {\bf 111}, no. 12, 121802 (2013)
  doi:10.1103/PhysRevLett.111.121802
  [arXiv:1212.4495 [hep-ph]].


\bibitem{Lester:1999tx} 
  C.~G.~Lester and D.~J.~Summers,
  Phys.\ Lett.\ B {\bf 463}, 99 (1999)
  doi:10.1016/S0370-2693(99)00945-4
  [hep-ph/9906349].

  
\bibitem{James:1975dr} 
  F.~James and M.~Roos,
  Comput.\ Phys.\ Commun.\  {\bf 10}, 343 (1975).
  doi:10.1016/0010-4655(75)90039-9
  

  
  \bibitem{Han:2009ss} 
  T.~Han, I.~W.~Kim and J.~Song,
  Phys.\ Lett.\ B {\bf 693}, 575 (2010)
  doi:10.1016/j.physletb.2010.09.010
  [arXiv:0906.5009 [hep-ph]].


\bibitem{Han:2012nr} 
  T.~Han, I.~W.~Kim and J.~Song,
  Phys.\ Rev.\ D {\bf 87}, no. 3, 035004 (2013)
  doi:10.1103/PhysRevD.87.035004
  [arXiv:1206.5641 [hep-ph]].
\bibitem{Han:2012nm} 
  T.~Han, I.~W.~Kim and J.~Song,
  Phys.\ Rev.\ D {\bf 87}, no. 3, 035003 (2013)
  doi:10.1103/PhysRevD.87.035003
  [arXiv:1206.5633 [hep-ph]].



\bibitem{Cho:2012er} 
  W.~S.~Cho, D.~Kim, K.~T.~Matchev and M.~Park,
  Phys.\ Rev.\ Lett.\  {\bf 112}, no. 21, 211801 (2014)
  doi:10.1103/PhysRevLett.112.211801
  [arXiv:1206.1546 [hep-ph]].
  
  
  
\bibitem{Alwall:2014hca} 
  J.~Alwall {\it et al.},
  JHEP {\bf 1407}, 079 (2014)
  doi:10.1007/JHEP07(2014)079
  [arXiv:1405.0301 [hep-ph]].
  
  
\bibitem{Ball:2013hta} 
  R.~D.~Ball {\it et al.} [NNPDF Collaboration],
  Nucl.\ Phys.\ B {\bf 877}, 290 (2013)
  doi:10.1016/j.nuclphysb.2013.10.010
  [arXiv:1308.0598 [hep-ph]].


\bibitem{Grigo:2014jma} 
  J.~Grigo, K.~Melnikov and M.~Steinhauser,
  Nucl.\ Phys.\ B {\bf 888}, 17 (2014)
  doi:10.1016/j.nuclphysb.2014.09.003
  [arXiv:1408.2422 [hep-ph]].
  


\bibitem{Czakon:2013goa} 
  M.~Czakon, P.~Fiedler and A.~Mitov,
  Phys.\ Rev.\ Lett.\  {\bf 110}, 252004 (2013)
  doi:10.1103/PhysRevLett.110.252004
  [arXiv:1303.6254 [hep-ph]].
  
\bibitem{Dittmaier:2011ti} 
  S.~Dittmaier {\it et al.} [LHC Higgs Cross Section Working Group],
  doi:10.5170/CERN-2011-002
  arXiv:1101.0593 [hep-ph].
  
\bibitem{deFlorian:2016spz} 
  D.~de Florian {\it et al.} [LHC Higgs Cross Section Working Group],
  doi:10.23731/CYRM-2017-002
  arXiv:1610.07922 [hep-ph].
  
\bibitem{deFlorian:2018wcj} 
  D.~de Florian, M.~Der and I.~Fabre,
  arXiv:1805.12214 [hep-ph].


\bibitem{Sjostrand:2006za} 
  T.~Sjostrand, S.~Mrenna and P.~Z.~Skands,
  JHEP {\bf 0605}, 026 (2006)
  doi:10.1088/1126-6708/2006/05/026
  [hep-ph/0603175].


\bibitem{Cacciari:2011ma} 
  M.~Cacciari, G.~P.~Salam and G.~Soyez,
  Eur.\ Phys.\ J.\ C {\bf 72}, 1896 (2012)
  doi:10.1140/epjc/s10052-012-1896-2
  [arXiv:1111.6097 [hep-ph]].
\bibitem{Cacciari:2008gp} 
  M.~Cacciari, G.~P.~Salam and G.~Soyez,
  JHEP {\bf 0804}, 063 (2008)
  doi:10.1088/1126-6708/2008/04/063
  [arXiv:0802.1189 [hep-ph]].
  
  \bibitem{ATL-PHYS-PUB-2013-004} 
  ATLAS Collaboration, 
  ATL-PHYS-PUB-2013-004.
  
   \bibitem{Sirunyan:2017ezt} 
  CMS Collaboration 
  JINST {\bf 13}, no. 05, P05011 (2018)

  
\bibitem{Kling:2016lay} 
  F.~Kling, T.~Plehn and P.~Schichtel,
  Phys.\ Rev.\ D {\bf 95}, no. 3, 035026 (2017)
  doi:10.1103/PhysRevD.95.035026
  [arXiv:1607.07441 [hep-ph]].

\bibitem{MIPtiming} 
L.~Gray and T.~Tabarelli de Fatis (Editors),
``Technical Proposal for a MIP Timing Detector in the CMS Experiment Phase 2 Upgrade",
CERN-LHCC-2017-027 / LHCC-P-009.

\bibitem{Asquith:2018igt} 
  L.~Asquith {\it et al.},
  arXiv:1803.06991 [hep-ex].

\bibitem{Soyez:2018opl} 
  G.~Soyez,
  arXiv:1801.09721 [hep-ph].
  
\bibitem{ATLAS:2012am} 
  G.~Aad {\it et al.} [ATLAS Collaboration],
  JHEP {\bf 1205}, 128 (2012)
  doi:10.1007/JHEP05(2012)128
  [arXiv:1203.4606 [hep-ex]].
\bibitem{Bertolini:2014bba} 
  D.~Bertolini, P.~Harris, M.~Low and N.~Tran,
  JHEP {\bf 1410}, 059 (2014)
  doi:10.1007/JHEP10(2014)059
  [arXiv:1407.6013 [hep-ph]].


\bibitem{Komiske:2017ubm} 
  P.~T.~Komiske, E.~M.~Metodiev, B.~Nachman and M.~D.~Schwartz,
  JHEP {\bf 1712}, 051 (2017)
  doi:10.1007/JHEP12(2017)051
  [arXiv:1707.08600 [hep-ph]].
  
  \bibitem{Martinez:2018fwc} 
  J.~Arjona Martínez, O.~Cerri, M.~Pierini, M.~Spiropulu and J.~R.~Vlimant,
  arXiv:1810.07988 [hep-ph].
  
  
   
\end{thebibliography}


\end{document}